\documentclass[letterpaper]{article}
\usepackage{aaai}
\usepackage{times}
\usepackage{helvet}
\usepackage{courier}
\usepackage{graphicx}
\usepackage{booktabs}
\usepackage{balance}
\newtheorem{definition}{Definition}

\begin{document}
%
\title{Improving Generalizability of Fake News Detection Methods using Propensity Score Matching}
 \author{Bo Ni,\\
{University of Notre Dame}\\
bni@nd.edu,
\And
Zhichun Guo,\\
{University of Notre Dame}\\
zguo5@nd.edu, 
\And
Jianing Li,\\
{University of Notre Dame}\\
jli23@nd.edu,
\And
Meng Jiang,\\
{University of Notre Dame}\\
mjiang2@nd.edu}
\maketitle

\begin{abstract}
\begin{quote}
Recently, due to the booming influence of online social networks, detecting fake news is drawing significant attention from both academic communities and general public. In this paper, we consider the existence of confounding variables in the features of fake news and use Propensity Score Matching (PSM) to select generalizable features in order to reduce the effects of the confounding variables. Experimental results show that the generalizability of fake news method is significantly better by using PSM than using raw frequency to select features. We investigate multiple types of fake news methods (classifiers) such as logistic regression, random forests, and support vector machines. We have consistent observations of performance improvement.
\end{quote}
\end{abstract}

\section{Introduction}
\label{introduction}
In recent years, due to the rapid development of online social networks, more and more people tend to seek out and obtain news from social media than from traditional media. While it certainly makes people's life richer and easier, it gives fake news a lot of chance to spread. Compared with traditional suspicious information such as email spam and web spam, fake news has much worse societal impact. First, fake news spreads faster and broader. Traditional suspicious information often targets specific recipients and only produces a local impact. However, online fake news can disseminate exponentially, affecting more people. Second, there is no or very little cost of creating suspicious contents on social media, which makes malicious users easier to create fake news. Indeed, fake news has contributed to a wide range of social problems such as the polarization between political parties.
 
Due to the negative impact of fake news, fake news detection has aroused world-wide interest. However, statistical causal inference, which indicates generalizable features of causal information to infer the identity of fake news, has not been extensively investigated in this area, leaving a couple of related issues unaddressed. We identify three challenges that need to be addressed.

\begin{itemize}
    \item First, there are various confounding variables in fake news corpora. Confounding variables are attributes that affect both dependent and independent variables. If they were left unnoticed when building machine learning algorithms, then we might not be able to learn generalizable causal features but only correlated features.
    
    \item
    Second, it is often difficult to extract useful features from text corpus. Traditional feature selection methods are susceptible to the existence of confounding variables. It is difficult to develop a feature selection method that can extract features of potential causal relationship. 
    
    \item
    Third, since traditional methods did not take causality into consideration, machine learning models that were trained on one dataset might suffer from significant performance depreciation encoutering a new dataset that has a slightly different data distribution. 
\end{itemize}

To address the above challenges, in this paper, we consider causal relations through a classical causality study method, Propensity Score Matching \cite{paul2017psm}. We obtain experimental data from open-source \texttt{FakeNewsNet} \cite{shu2018fakenewsnet}, which consists of data from two different sources, \texttt{politifact} and \texttt{gossipcop}. Then we select word features using Propensity Score Matching proposed. We adopted logistic regression in propensity score calculation. We evaluate feature selection methods using logistic regression and also extend them to other machine learning models. We conduct comparative analysis between datasets. Our method successfully improves the generalizability of the classifiers across multiple datasets.
 


\subsection{Reproducibility}

The \texttt{FakeNewsNet} \cite{shu2018fakenewsnet} that we used in our paper is publicly available and can be found online.\footnote{https://github.com/KaiDMML/FakeNewsNet}
Our code is publicly available.\footnote{https://github.com/Arstanley/fakenews\_pscore\_match/}

\section{Related work}
\label{related work}
In this section, we discuss relevant topics to our proposed research work. The topics mainly include fake news detection and causal inference.

\subsection{Fake News Detection}

Online fake news detection has attracted a lot of attention from researchers. Most of them focus on applying machine learning classifiers to automatically identify fake news. We will summarize them into four categories based on the types of features they extracted and used.

\begin{itemize}
\item \textbf{Linguistic features extracted from news.}
Castillo \emph{et al.} used a series of linguistic features from news such as content length, emoticon, hashtag, etc. to access the credibility of a given set of tweets \cite{castillo2011information}. Swear words, emotion words and pronouns are extracted to do credibility assessment \cite{gupta2014tweetcred}. Moreover, assertive verbs and factive verbs have been used \cite{potthast2017stylometric}.

\item \textbf{Linguistic features extracted from user comments.} User comments can reflect the authenticity of news. Zhao \emph{et al.} detect fake news by inquiry phrases from users comments \cite{zhao2015enquiring}. Ma \emph{et al.} and Chen \emph{et al.} used RNN-based methods which captured linguistic features from users comments to detect rumors \cite{ma2016detecting,chen2018call}.    

\item \textbf{Structure features extracted from social networks.} Wu \emph{et al.} proposed a graph kernel based
hybrid SVM classifier which captured the high-order propagation patterns in addition to semantic features to do fake news detection \cite{wu2015false}. Sampson \emph{et al.} classified conversations through the discovery of implicit linkages between conversation fragments \cite{sampson2016leveraging}.

\item \textbf{Combine different types of features.} Castillo \emph{et al.} usedfeatures from user's posting and re-posting behavior, from the text of the
posts, and from citations to external sources \cite{castillo2011information}. Yang \emph{et al.} combined content-based, user-based, client-based, and location-based features \cite{yang2012automatic}. Kwon \emph{et al.} examined a comprehensive set of user, structural, linguistic, and temporal features \cite{kwon2017rumor}. 
\end{itemize}

Despite traditional methods of fake news detection, recently researchers focus on more specific yet challenging problems in this domain. Zhang \emph{et al.} detected fauxtography (misleading images) on social media using directed acyclic graphs produced by user interactions \cite{zhang2018fauxbuster}. Wang \emph{et al.} adopted adversarial neural networks for early stage fake news detection \cite{wang2018eann}. Shu \emph{et al.} employed a co-attention neural network to detect fake news in an explainable framework using both user-based features and content-based features \cite{shu2019defend}.

In this paper, our work will focus on the generalizing ability of machine learning models. The main objective of our work is not to improve the absolute performance of fake news detection. Here we employ content-based features only but our proposed methods can be generalized to other types of features if causal relationship exists.

\subsection{Causal Machine Learning}

Most of existing popular machine learning algorithms hold the assumption of independent and identically distributed (IID) data. Indeed they have reached impressive results in various big data problems \cite{lecun2015learning}. However, this is a strong assumption and not suitable for a lot of real world situations \cite{scholkopf2019causal}. Recently, various researchers have achieved great progress in causal machine learning. Pearl \emph{et al.} introduced the causal graphs and structural causal models, incorporating the notion of intervention into statistical machine learning models \cite{pearl2009causality}. In addition, confounding variables have been discussed in different domains. Lu \emph{et al.} addressed the presence of confounding variable in the setting of reinforcement learning by extending an actor-critic reinforcement learning algorithm to its de-confounding variant \cite{lu2018deconfounding}. In the realm of text classification, Landeiro studied the problem by explicitly indicating the specific confounding variables that might misguide the classifier \cite{landeiro2016confounding}. 

Propensity score matching, the technique that we use in this work, was proposed by Rosenbaum \emph{et al.} to address the presence of confounding variables in statistical experiments \cite{rosenbaum1985psm}. It has been extended to user-generated data and sentiment classification \cite{rehman2016psm,dos2015using,paul2017psm}. Paul \emph{et al.} generalized the propensity score matching to a feature selection technique that takes confounding variables into consideration \cite{paul2017psm}. Causal methods have not been fully investigated in the realm of fake news detection.

\section{Problem Definition}
\label{problem formulation}

In this section, we formally define \textit{confounding variables} and the task of \textit{de-counfounding fake news detection}.

\begin{definition}[Confounding Variables]
Let $X$ be some independent variable, $Y$ some dependent variable. We say $Z$ is a confounding variable that confounds $X$ and $Y$ if $Z$ is an unobserved variable that influence both X and Y \cite{wikiconfounding}.
\end{definition}

In this paper, we focus on mitigating the effects of confounding variables in fake news detection. First, we explain why causal techniques are necessary, and we justify the presence of confounding variables in fake news detection. Indeed, a great amount of fake news have political purpose. As a result, one classifier might take word ``trump'' as a useful feature, but intuitively, ``trump'' does not \textit{causally} indicate a piece of news being fake. To make predictions based on such features will inevitably result in weak robustness. When encountering another news dataset that might be less political, it would perform worse. In order to address this issue, We formulate the problem as follows:

Let $N=(T, C)$ be a news entity that consists of a title $T$ and contents $C$, $L=\{Fake, Real\}$ a binary label, we define the problem of de-counfounding fake news detection as follows: Assume there exists a confounding variable Z that confounds $N$ and $L$. Given a dataset $S=\{(N_i, L_i) \mid 1<=i<=n\}$ which consists of all the news and their corresponding labels, we aim at learning a map from the input space $N$ to the label space $L$.

\section{Proposed Approach}
\label{methodology}
\subsection{Overview}

In this work, we use propensity score matching to select \textit{decounfounded} features for fake news detection. Following the methods proposed in \cite{paul2017psm}, for each feature, we first calculate the propensity score of every sample regarding the specific feature. Then, we employ a one-to-one matching based on the propensity score. Finally, chi-square test statistics are used to rank the causal relevance of the features. More details are provided in the following sections.

\subsection{Propensity Score}
In statistics, causality analysis is often conducted with control-treatment pairs. However, in most of the real-word situations, it is impossible to obtain control-treatment pairs. Propensity score intends to solve the problem. As initially proposed in \cite{rosenbaum1985psm}, it is defined as the probability of a subject to receive a certain treatment. In the realm of fake news detection, we regard each word feature as a treatment and each news sample as a subject. Then we formally define the propensity score as below.

\begin{definition}[Propensity Score]
Let $w$ be a word feature, $X$ a news corpus. We say $psm(w, X)$ is the propensity score of $w$ regarding $X$ and 
$$psm(w, X) = P(w|X-\{w\})$$
\end{definition}

The propensity score can be estimated in many different ways \cite{austin2011psm}. In this work, we conduct experiment with logistic regression and random forest regression. And we compare them in the experiment section.

\subsection{Matching}

By pairing the subjects that have similar propensity scores, we can eliminate the bias caused by confounding variables. Among different matching strategies \cite{austin2011psm}, we use one-to-one matching for its efficiency. We rank the subjects by their propensity score and greedily find the matched subjects. Each pair of matched subjects consists of (1) a treatment unit, (2) a text corpus that contains the word feature and a control unit, and (3) a text corpus that does not contain the word feature but has a similar propensity score. We finally calculate the chi-square test statistics for each feature with the paired subjects with
\begin{equation}
X^2 = \frac{(TN-CP)^2}{TN+CP},
\end{equation}
where $TN$ stands for treatment-negative and $CP$ stands for control-positive.

\section{Experiments}
\label{Experiments}
\subsection{Dataset}

We conduct experiments to evaluate our propensity score matching-based approach with a widely used fake news dataset, \texttt{FakeNewsNet} \cite{shu2018fakenewsnet}. The dataset consists of news corpora from two primary sources:

\begin{itemize}
    \item \texttt{PolitiFact}: In \texttt{PolitiFact}, political news collected from various sources are fact-checked by experts and journalists. Specifically, we use the sample data provided by \cite{shu2018fakenewsnet}. It consists of $1,056$ data points for the \texttt{PolitiFact} section. And it includes 624 real news and 432 fake news documents.
    \item \texttt{GossipCop}: \texttt{GossipCop} is a website that collects entertainment stories from various sources with fact-check score that ranges from 1 to 10. However, since the website intends to showcase mostly fake stories, the majority of the stories have scores less than five. Real stories are collected from \textit{E! Online}, a widely-accepted reliable entertainment website. The samples that we use in this work consists of 16,817 real stories and 5,323 fake stories. 
\end{itemize}

Since for both \texttt{PolitiFact} and \texttt{GossipCop}, real samples are fewer than fake samples, in order to balance the label distribution, we randomly sample 432 and 5323 real samples from the two datasets, respectively.

\subsection{Experiment Settings and Results}

\begin{figure}[t]
    \centering
    \includegraphics[width=0.5\textwidth]{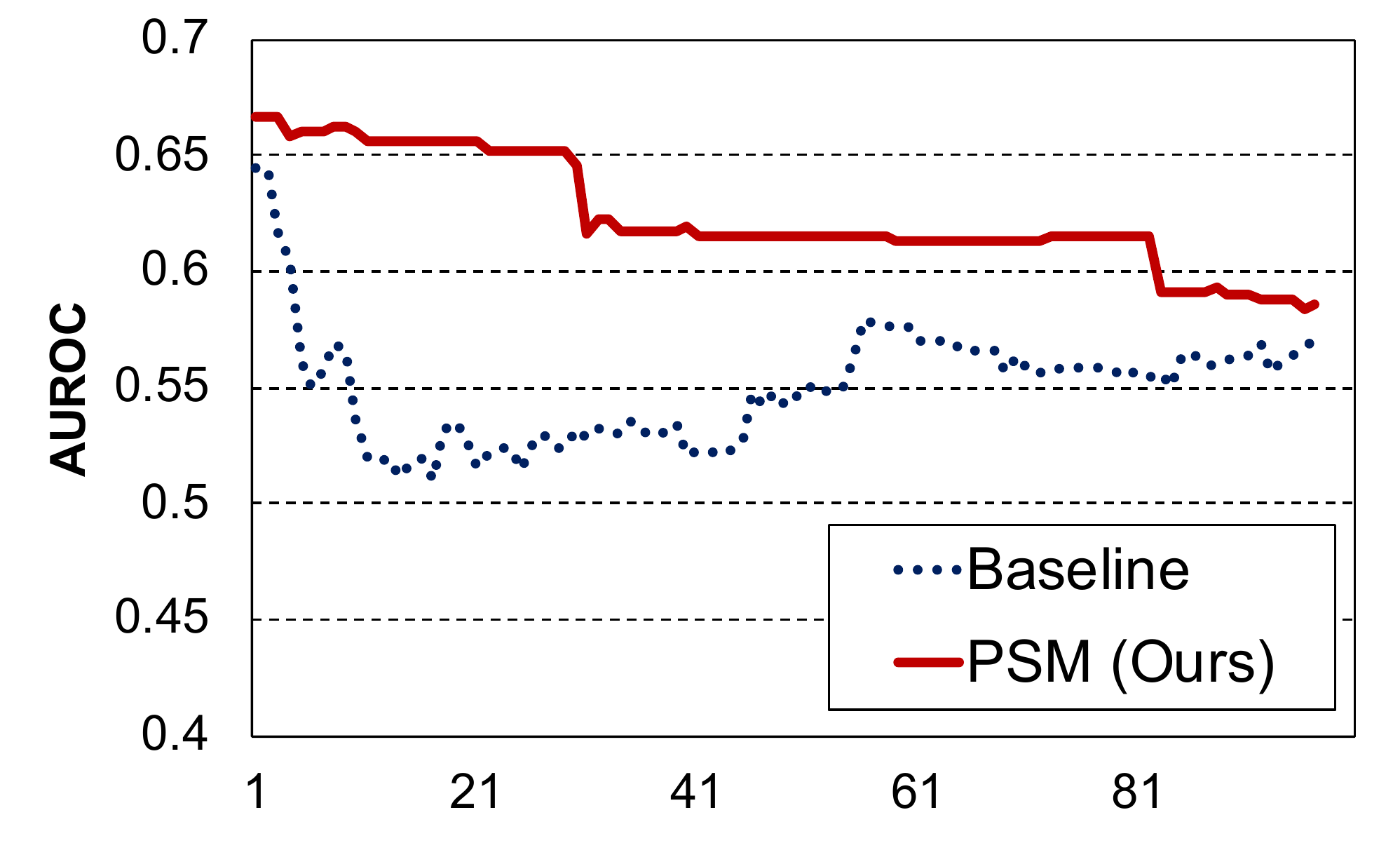}
    \caption{We developed a fake news classification model and trained it  on \texttt{PolitiFact}. We evaluated the model on \texttt{GossipCop}. Clearly, our PSM-based method performs better than the baseline method (higher AUROC).}
    \label{fig:log-clf-politifact}
\end{figure}

\begin{figure}[t]
    \centering
    \includegraphics[width=0.5\textwidth]{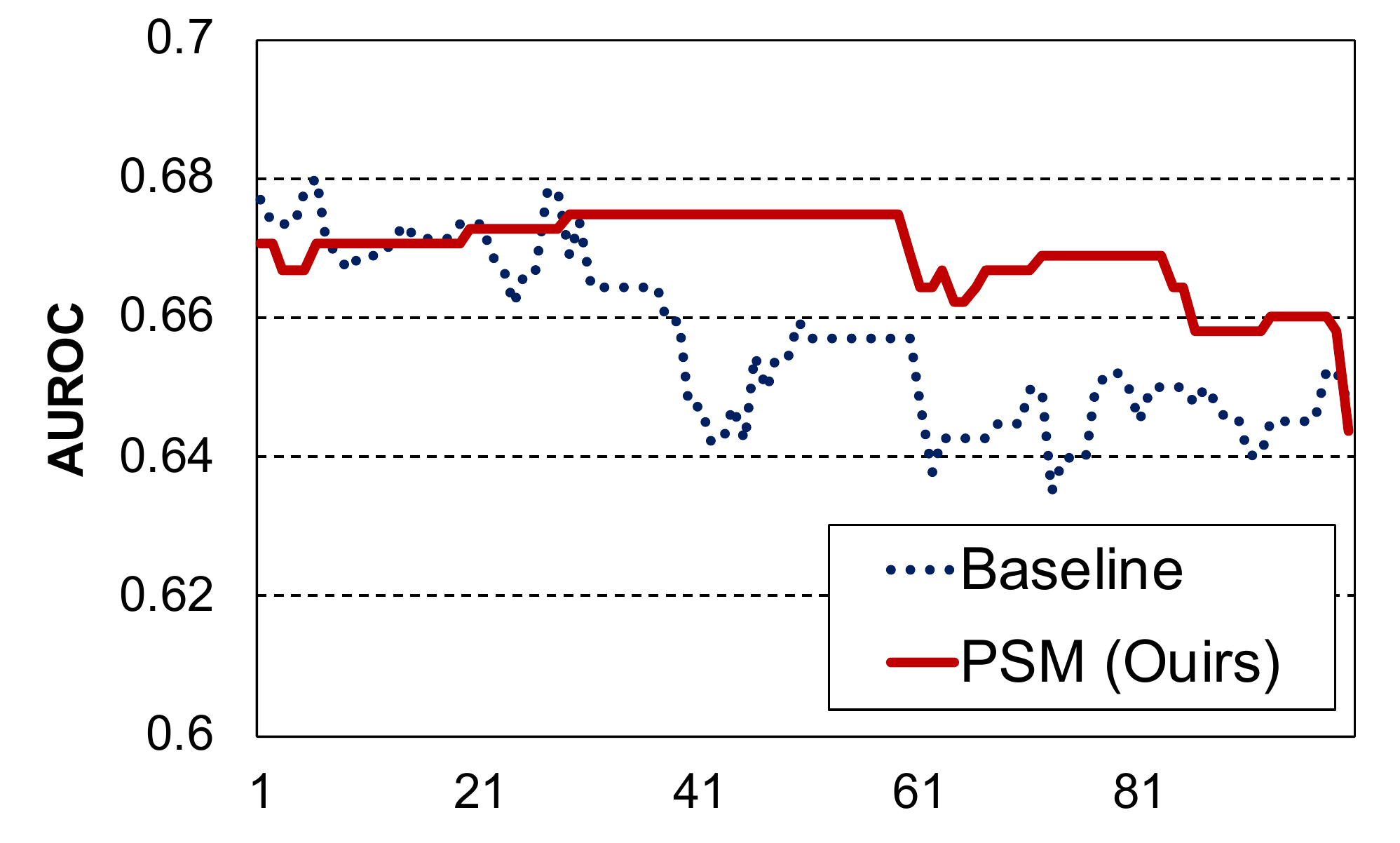} 
    \caption{We developed a fake news classification model and trained it  on \texttt{GossipCop}. We evaluated the model on \texttt{PolitiFact}. Clearly, our PSM-based method performs better than the baseline method (higher AUROC).}
    \label{fig:log-clf-gossipcop}
\end{figure}

We use document frequency as the baseline feature selection method and we intend to compare the generalization ability. Also, we want to observe how well the model performance evolves when adding more features. For simplicity, we consider a standard logistic classifier with default parameters provided by scikit-learn \cite{scikit-learn}. We train two classifiers on both datasets separately with different percentage of features and then evaluate on each other. We visualize our results in Figure \ref{fig:log-clf-politifact} and Figure \ref{fig:log-clf-gossipcop}. The curve starts at the point where 1\% of the top features are selected. As we can see, PSM consistently outperforms the baseline feature selection method based on document frequency in the task of fake news prediction across data regardless of the percentage of features used. We notice that the gap is smaller in Figure \ref{fig:log-clf-gossipcop} than that in Figure \ref{fig:log-clf-politifact}. It is due to the fact that entertainment stories have more diverse themes which makes feature selected by document frequency more reliable. In general, as illustrated by both graphs, by applying propensity score matching to feature selection, we obtain models with better generalization ability. To summarize the graph to a more concise metric, we calculate the AUROC in Table \ref{tab:auroc}. 

It is noticeable that when using a relatively small percentage of selected features, our PSM-based method significantly outperforms the baseline method. It is reasonable since the top features selected from PSM should be more accurately representative of what fake news might look like and less subject to specific topics. To illustrate this point, we provide an empirical analysis in the next section.

\subsection{Empirical Analysis}

To better understand how propensity score matching improves the generalization ability, we showcase some specific examples in this section. We only consider the model trained on \texttt{PolitiFact} in this section for a clearer representation. We show top five features from the baseline method and also top five features from propensity score matching in Table \ref{tab:words}. Clearly, when using document frequency for feature selection, most of the top features turn to be political figures. It is self-evident that features selected by document frequency only reflects the distribution of the \texttt{PolitiFact} dataset, so it cannot be generalized well when there is a change of distribution. In contrast, top features acquired from PSM are more likely to be patterns of fake stories in general, and words like ``confirmed'' and ``inside'' could be reliable features that generalize to other datasets. 

\begin{table}[t]
    \centering
    \begin{tabular}{|c|c|c|}
    \hline
     & Baseline & PSM (Ours) \\
    \hline \hline
     \texttt{PolitiFact} &  0.56 & \textbf{0.68}   \\
    \hline
     \texttt{GossiCop}   &  0.63 & \textbf{0.67}   \\
    \hline
    \end{tabular}
    \caption{Experimental results (AUROC Score) show that PSM-based method performs better than baseline method.}
    \label{tab:auroc}
\end{table}

\begin{table}[t]
    \centering
    \begin{tabular}{c|c}
    \hline
     Baseline & PSM (ours) \\
    \hline \hline
      ``trump''  & ``makes'' \\
      ``obama''  & ``leaves'' \\
      ``senator'' & ``confirmed'' \\
      ``donald'' & ``nightmare'' \\
      ``action'' & ``inside'' \\
    \hline
    \end{tabular}
    \caption{Top features that the baseline method and our PSM-based method discovered in the \texttt{PolitiFact} dataset.}
    \label{tab:words}
\end{table}

\section{Future Work}
\label{futurework}
Although we obtain improvements in the generalization of fake-news detection, there still remains a couple of challenges in this area. One thing worth noticing is that PSM only accounts for biases caused by observed variables. Researchers could focus on mitigating the biases caused by latent variables. One approach could be extending PSM to latent representations learned by deep neural networks. Another direction of improvements could be causal fake-news detection with Bayesian Networks and Structural Equation Models proposed by Pearl \emph{et al.} \cite{pearl2009causality} which ensures there will be no hidden confounding variables. 

\section{Conclusions}
\label{conclusion}
In this work, we conducted a data-driven study on the generalization ability of fake news detection models. We approached the task by introducing propensity score matching into the feature selection process. We study the generalization ability of fake news detection models. In conclusion, our experimentation shows significant improvement of using propensity score matching as feature selection compared with baseline model on the generalizability.

\clearpage
\balance

\bibliographystyle{aaai}
\bibliography{reference,jiang}
\end{document}